\documentclass{SciPost_better_arXiv}

\hypersetup{
    colorlinks,
    linkcolor={red!50!black},
    citecolor={blue!50!black},
    urlcolor={blue!80!black}
}

\usepackage[bitstream-charter]{mathdesign}
\usepackage{booktabs}
\usepackage[export]{adjustbox}
\usepackage{subcaption}

\DeclareSymbolFont{usualmathcal}{OMS}{cmsy}{m}{n}
\DeclareSymbolFontAlphabet{\mathcal}{usualmathcal}

\fancypagestyle{SPstyle}{
\fancyhf{}
\lhead{\colorbox{scipostblue}{\bf \color{white} ~SciPost Physics Community Reports }}
\rhead{{\bf \color{scipostdeepblue} ~Submission }}

\fancyfoot[C]{\textbf{\thepage}}
}

\RequirePackage{heppennames2}
\RequirePackage{lhchiggs}

\newcommand{\GeV}{\mathrm{GeV}}

\usepackage{comment}

\usepackage[dvipsnames]{xcolor}

\begin{document}

\pagestyle{SPstyle}
\begin{flushright}\footnotesize{
CERN-TH-2025-235}\end{flushright}
\begin{center}{\Large \textbf{\color{scipostdeepblue}{
Interference effects in gluon-fusion Higgs boson production
}}}\end{center}

\begin{center}\textbf{
Federico Buccioni\textsuperscript{1,2},
Federica Devoto\textsuperscript{3}
}\end{center}

\begin{center}
{\bf 1} Technische Universit\"at M\"unchen,\\Physik Department,
James-Franck-Straße 1,
D–85748 Garching, Germany
\\
{\bf 2} CERN, Theoretical Physics Department, 1211, Geneva 23, Switzerland
\\
{\bf 3} SLAC National Accelerator Laboratory, Stanford University, Stanford, CA 94039, USA
\\[\baselineskip]
\href{mailto:federico.buccioni@cern.ch}{\small federico.buccioni@cern.ch},
\href{mailto:federica@slac.stanford.edu}{\small federica@slac.stanford.edu}, 
\end{center}

\section*{\color{scipostdeepblue}{Abstract}}
\textbf{\boldmath{%
In this contribution we summarize recent progress in higher-order computations of signal-background interference effects in Higgs boson production via gluon fusion. The focus is on the resonance region, where interference terms are most significant relative to the pure signal contribution when the Higgs boson decay is loop-induced. We present results for the well-studied $gg \to H \to \gamma\gamma$ process and for the rare $gg \to H \to Z\gamma$ mode. In both cases, the interference is destructive, reducing the resonant Higgs boson production rate by about 1.6\% in the diphoton channel and by roughly 3\% in the $Z\gamma$ decay mode.
}}

\vspace{\baselineskip}

\noindent\textcolor{white!90!black}{%
\fbox{\parbox{0.975\linewidth}{%
\textcolor{white!40!black}{\begin{tabular}{lr}%
  \begin{minipage}{0.6\textwidth}%
    {\small Copyright attribution to authors. \newline
    This work is a submission to SciPost Phys. Comm. Rep. \newline
    License information to appear upon publication. \newline
    Publication information to appear upon publication.}
  \end{minipage} & \begin{minipage}{0.4\textwidth}
    {\small Received Date \newline Accepted Date \newline Published Date}%
  \end{minipage}
\end{tabular}}
}}
}\newpage


\vspace{10pt}
\noindent\rule{\textwidth}{1pt}
\tableofcontents
\noindent\rule{\textwidth}{1pt}
\vspace{10pt}

\section{Introduction}
At the Large Hadron Collider (LHC)
the dominant production mechanism for a single Higgs boson
is through gluon–gluon scattering. In this partonic channel, 
the Higgs boson couples to a loop of heavy quarks excited by the initial-state gluons.
In the Standard Model (SM), owing to their strong Yukawa interaction, top quarks running
in the loop provide the largest contribution, followed by bottom quarks, whose interaction
strength with the Higgs boson is roughly 50 times smaller than that of the top quarks.

Given the importance of this production channel, considerable effort has been devoted to providing theoretical predictions for the cross section with the best possible accuracy.
Indeed, it is well known that gluon-fusion Higgs production suffers from large perturbative corrections in Quantum Chromodynamics (QCD)\cite{Georgi:1977gs,Graudenz:1992pv}. This feature calls for the inclusion of higher-order effects to ensure reliable predictions and observe first signs of convergence of the perturbative series. Working in the infinite top-quark mass limit, next-to-next-to-leading order (NNLO) QCD corrections were computed more than twenty years ago~\cite{Harlander:2002wh,Anastasiou:2002yz,Ravindran:2003um}, while third-order ($\mathrm{N^3LO}$) ones became available ten years ago~\cite{Anastasiou:2015vya,Anastasiou:2016cez,Mistlberger:2018etf}.
Very recently, first results beyond $\mathrm{N^3LO}$ accuracy have been presented in the literature~\cite{Das:2025tuk}.

Besides large perturbative QCD corrections, a broad range of effects 
contributing to the theoretical uncertainty budget
were identified in Ref.~\cite{LHCHiggsCrossSectionWorkingGroup:2016ypw}. 
Many such effects individually impact the theoretical accuracy at the 1\% level, and collectively amount to several percents. 
Some of the dominant sources have been recently addressed in the 
literature. They include mixed QCD-electroweak corrections~\cite{Bonetti:2017ovy,Anastasiou:2018adr,Bonetti:2018ukf,Becchetti:2020wof},
finite top-mass effects~\cite{Czakon:2021yub}, and heavy-light quark interference
contributions~\cite{Czakon:2023kqm}.
Currently, the two outstanding sources of uncertainty come from the strong coupling constant $\alpha_s$
and the lack of robust $\mathrm{N^3LO}$ PDF fits, see e.g.~\cite{Cridge:2024icl} as well as other 
contributions to this report.

The discussion so far concerned only the gluon-fusion production stage. 
However, when the fully decayed process is considered, 
other types of effects come into play. 
An interesting one is the quantum interference 
between resonant Higgs production and its subsequent decay, 
the signal process, and the continuum background~\cite{Dicus:1987fk,Dixon:2003yb}.
The latter features the same initial- and final-state
configurations, but is not mediated by a Higgs boson.
As we will discuss later, interference effects on the on-shell cross section are larger when the Higgs boson decay is loop induced, such as in
the $H\to\gamma\gamma$ and $H\to Z\gamma$ modes.
Although next-to-leading order (NLO) QCD predictions in the diphoton case were presented long ago~\cite{Dixon:2003yb,Dixon:2013haa,Campbell:2017rke}, 
this type of contribution was never accounted for in the theory uncertainty budget, 
despite its impact being comparable to other more studied effects, such as the ones mentioned above. 
Thanks to the advances in perturbative QCD calculations, 
signal-background interference in on-shell Higgs production 
has recently attracted some renewed attention and predictions have been improved beyond the previous state of the art~\cite{Bargiela:2022dla,Buccioni:2023qnt}. 
The impact of such effects on the on-shell rate has been calculated to be above one percent level both in the diphoton channel and the rare $Z\gamma$ decay.

In the next section we will discuss the main features of signal-background interference effects in the case of on-shell Higgs boson production. 
In Section~\ref{sec:gammagamma} we present results in the case of the $H\to\gamma\gamma$ decay, whereas in Section~\ref{sec:zgamma} those for $H\to Z\gamma$.
We conclude and summarize the main findings in Section~\ref{sec:conclusions}.

\section{Interference effects in Higgs mediated diboson production}
\label{se:interferencevv}
In this note, we are concerned with signal-background interference contributions to on-shell Higgs boson production. 
In particular, we focus on the dominant gluon-fusion channel, followed by the Higgs boson decay to a pair of vector bosons $(V_1 V_2)$.
In principle, interference effects are present in all diboson Higgs decays, i.e. $ZZ$, $WW$, $\gamma\gamma$ and $Z\gamma$. 
However, the impact of such effects relative to the signal rate
are larger when the Higgs boson decay is loop induced (often referred to as \textit{loop enhancement} in the literature).
This is easily understood since the signal amplitude itself is suppressed by a loop factor $\mathcal{O}(\frac{\alpha}{4\pi})$ with respect to tree-level decays such as $H\to Z^*Z$ and $H\to W^{+}W^{-}$
\footnote{We point out that in off-shell Higgs production instead, interference effects in the $ZZ$ and $W^{+}W^{-}$ signatures are non-negligible and have been studied in the literature, see e.g.~\cite{Bonvini:2013jha,Caola:2016trd}}.
Therefore, from now on, we focus on the decay modes $H\to\gamma\gamma$ and $H\to Z\gamma$, where the interference effects around the Higgs resonance are the largest.

We start by writing the scattering amplitude for the process $gg \to V_1 V_2$ as the sum of the Higgs-mediated contribution and the continuum background as follows
\begin{equation}
\label{eq:amplitude}
    \mathcal{M}_{gg\to V_1 V_2} = \frac{\mathcal{M}_{\rm sig}}{m_{V_1 V_2}^2 - m_H^2 + i \Gamma_H m_H} + \mathcal{M}_{\rm bkg}\, ,
\end{equation}
where in the signal amplitude we factored out the Breit-Wigner propagator for convenience.
If we consider the invariant mass distribution of the diboson system (also referred to as the line shape), this can be schematically written as
\begin{equation}
    \frac{\rm{d}\sigma}{{\rm{d}} m^2_{V_1 V_2}} \propto 
    \frac{|\mathcal{M_\mathrm{sig}}|^2}{(m^2_{V_1 V_2}-m_H^2)^2 + \Gamma_H^2 m_H^2} + |\mathcal{M}_\mathrm{bkg}|^2 + I,
\end{equation}
where the interference term $I$ is given by
\begin{equation}
    I = 2\, {\rm Re}\left(\frac{\mathcal{M}_{\rm sig}}{m_{V_1 V_2}^2 - m_H^2 + i \Gamma_H m_H} \mathcal{M}_{\rm bkg}^\dagger\right)\,.
\end{equation}
If one further decomposes the signal and background amplitudes into their real and imaginary parts, the interference contribution can be expressed as a sum of two different terms,
\begin{equation}
\label{eq:reimparts}
\begin{split}
    I_{\rm Re} = & \frac{2 \,(m_{V_1 V_2}^2-m_H^2)}{(m_{V_1 V_2}^2-m_H^2)^2 +\Gamma_H^2 m_H^2}
    \left[{\rm Re}\mathcal{M}_{\rm sig}{\rm Re}\mathcal{M}_{\rm bkg} + {\rm Im}\mathcal{M}_{\rm sig} {\rm Im}\mathcal{M}_{\rm bkg}\right] \,, \\
     I_{\rm Im} = & \frac{2 \,\Gamma_H m_H}{(m_{V_1 V_2}^2-m_H^2)^2 +\Gamma_H^2 m_H^2} 
    \left[{\rm Re}\mathcal{M}_{\rm bkg}{\rm Im}\mathcal{M}_{\rm sig} - {\rm Im}\mathcal{M}_{\rm bkg} {\rm Re}\mathcal{M}_{\rm sig}\right]\,.
\end{split}
\end{equation}
We are going to denote them as the \textit{real} and \textit{imaginary} part of the interference respectively.\footnote{Note that both these quantities are strictly real numbers, and the naming \textit{real} and \textit{imaginary} just follows from the relevant literature.}. 
\begin{figure}[t!]
    \centering
    \includegraphics[width=0.67\linewidth]{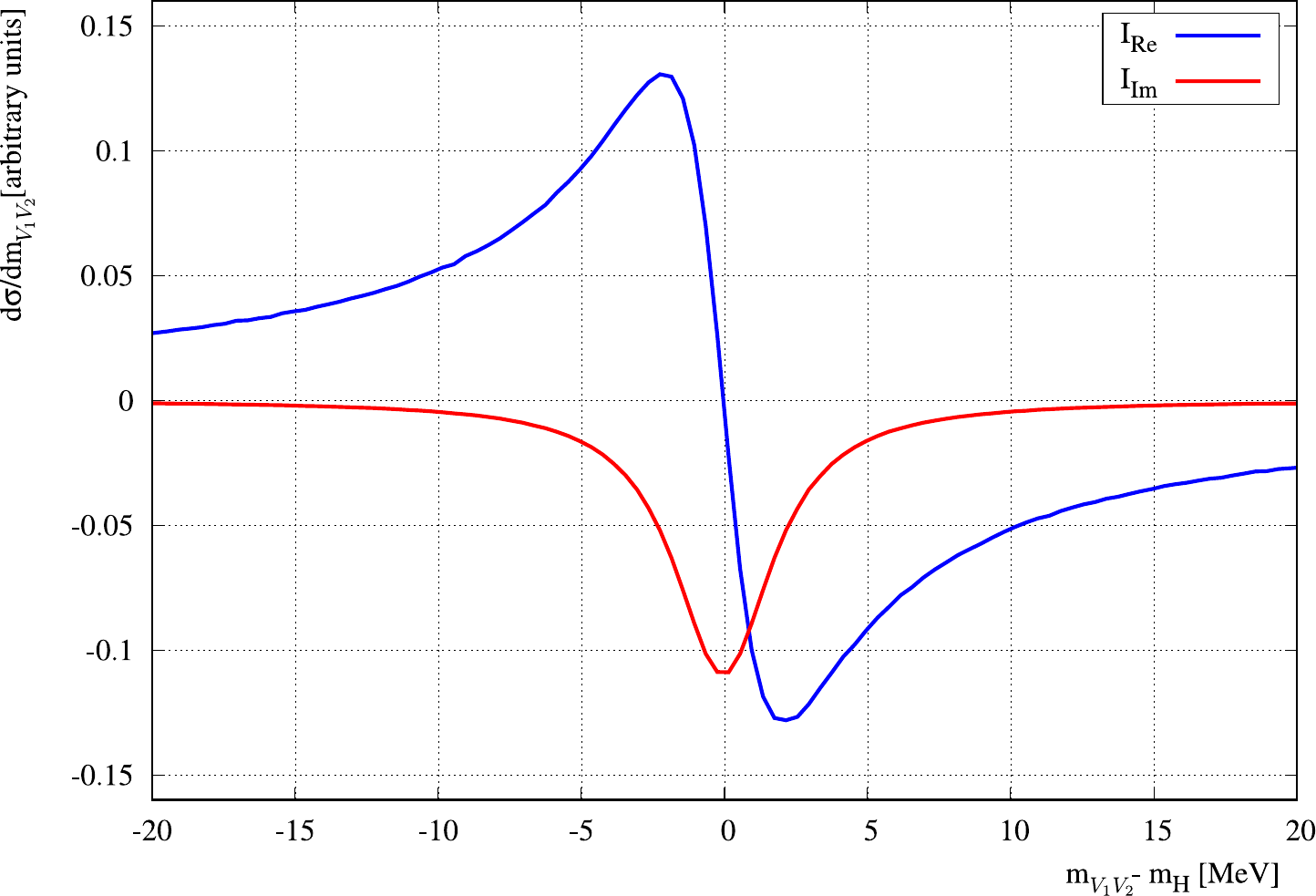}
    \caption{Schematic representation of the invariant mass distribution for the real (blue) and imaginary (red) contributions to interference as defined in Eq.~\ref{eq:reimparts}.}
    \label{fig:reimpart}
\end{figure}
These two contributions have clearly distinct features.
For illustrative purposes only, we show their line shapes separately in Fig.~(\ref{fig:reimpart}).

The real part, $I_{\rm Re}$, is antisymmetric around the Higgs resonance, thus its contribution to the on-shell Higgs cross section is negligible. Its prominent effect is to generate an imbalance of events around the resonance, effectively
shifting the location of the Higgs peak~\cite{Martin:2012xc}.
In Ref.~\cite{Dixon:2013haa} it was pointed out how this \textit{mass shift} effectively lifts a degeneracy in a simultaneous extraction of couplings and Higgs-boson total decay width ($\Gamma_H$) in on-shell cross sections measurements. 
In particular,
given the highly accurate measurements in the diphoton channel, if the mass shift is extracted with sufficient precision, a comparison against accurate theoretical predictions could provide valuable bounds on $\Gamma_H$~\cite{Dixon:2013haa,Bargiela:2022dla}. For a detailed discussion on the role of theory predictions on such bounds, we refer the interested reader to~\cite{Dixon:2013haa,Bargiela:2022dla}.
Since the original proposal~\cite{Dixon:2013haa},
further refinements at NLO QCD accuracy  were discussed in~\cite{deFlorian:2013psa,Coradeschi:2015tna,Cieri:2017kpq},
whereas first results beyond NLO QCD accuracy were presented in~\cite{Bargiela:2022dla}.
On the experimental front, 
the ATLAS collaboration explored the feasibility of this approach in~\cite{ATLAS:2016kvj} reporting looser bounds compared to the ones estimated from theory.
The CMS collaboration instead exploited the actual distortion in the lineshape to significantly enhance sensitivity to these effects~\cite{CMS:2025zue}, achieving a level of precision comparable to that of theoretical predictions~\cite{Dixon:2013haa}.
Thus, bounding $\Gamma_H$ via the mass-shift based strategy is not as competitive as the off-shell strategy yet~\cite{Caola:2013yja,CMS:2024eka,ATLAS:2025okx}.
However, since it relies on an on-shell cross section measurement, it has minimal model dependence, thus
providing valuable complementary information.

\begin{figure}[t!]
    \centering
    \includegraphics[width=1.\linewidth]{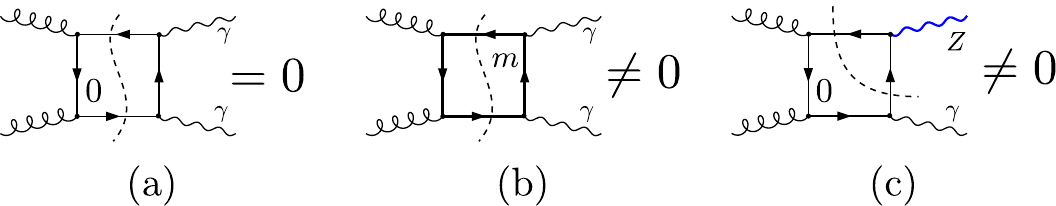}
    \caption{Imaginary part of one-loop background amplitudes. The dashed line represents a unitarity cut. In (a) and (b) it corresponds to an $s$-channel cut whereas in (c) to an $m^2_Z$ cut. Thick and thin lines in the loops represent massive and massless quarks respectively.
    In (a) and (b) both photons are assumed to have the same helicity. In order for (b) to develop an imaginary part one needs to have $\sqrt{s}=m_H>2m$.}
    \label{fig:unitaritycuts}
\end{figure}
The imaginary part $I_{\rm Im}$, instead, is symmetric around the resonance, therefore it contributes to the total cross section. In the SM, such interference turns out to be destructive~\cite{Dixon:2003yb,Campbell:2017rke,Bargiela:2022dla}.
The impact of the imaginary part relative to the on-shell Higgs cross section can be simply estimated by integrating the Breit-Wigner propagator in the second row of Eq.~\eqref{eq:reimparts} over $m^2_{V_1 V_2}$. Taking into account coupling factors, one gets
\begin{equation}
\label{eq:interferenceestimate}
    \bigg|\frac{\sigma_{\rm{int}}}{\sigma_{\rm{sig}}}\bigg|\simeq \frac{2\Gamma_H}{m_H}\frac{(4\pi v)^2}{m_H^2} \sim 4\%\,,
\end{equation}
where $v$ is Higgs-field vacuum expectation value.
However, this estimate can vary substantially depending on the Higgs-boson decay mode. As for the $\gamma\gamma$ channel,
given the scalar nature of the Higgs boson, both photons have opposite spin, thus identical helicity. In this scenario, at LO, the 
background amplitude in Eq.~\eqref{eq:amplitude} develops an imaginary contribution only in the presence of massive quarks in the loops, and if the two-particle threshold is below the Higgs mass, see Fig.~(\ref{fig:unitaritycuts},a-b). Thus,
the largest contribution comes from a loop of bottom quarks, but its scaling behaviour is $m_b^2/m_H^2$. 
It then follows from the second row of Eq.~\eqref{eq:reimparts}
that $I_{\mathrm{Im}}$ is mass suppressed, impacting the cross-section well below the percent level. The picture drastically changes 
at NLO because large imaginary terms appear in the two-loop background amplitudes from diagrams involving massless quarks~\cite{Dixon:2003yb,Campbell:2017rke}. 
The destructive interference reduces the NLO on-shell Higgs cross section by more than $1\%$~\cite{Dixon:2003yb,Campbell:2017rke,Bargiela:2022dla}. 
Although this appears at the two-loop level, it is effectively a LO effect, which strongly motivated investigations through higher orders. We provide a detailed account of the impact of higher-order corrections in Section~\ref{sec:gammagamma}.

As for the $Z\gamma$ decay mode, the qualitative features of the real and imaginary parts are similar to the $\gamma\gamma$ case. Namely, the real part of the interference is responsible for a mass-peak shift, and the imaginary part provides a destructive contribution to the cross section.
However, contrary to the diphoton case, there is no helicity selection rule on the final-state products. Moreover, the presence of a massive $Z$ boson, which couples to the quarks in the loop, provides a non-vanishing unitarity cut. Thus,
unlike in the $\gamma\gamma$ case, loops of light quarks provide a destructive interference contribution already at LO~\cite{Buccioni:2023qnt}, see Fig.~(\ref{fig:unitaritycuts}c).
Finally, it is worth noting that in the $Z\gamma$ case, the real part of the interference has the opposite sign 
with respect to the diphoton case, thus generating an excess of events to the right of Higgs boson peak~\cite{Buccioni:2023qnt}.

\section{The $H\to\gamma\gamma$ decay channel}
\label{sec:gammagamma}
In this section we present results for interference effects in the $\gamma \gamma$ decay mode beyond NLO QCD accuracy. 
We already argued in the previous section why the effects on the total cross section are negligible at LO, and can be appreciated only starting from NLO. This motivates the need to control such contributions to even higher orders in perturbation theory. 

The NLO QCD analysis was performed in Ref.~\cite{Dixon:2003yb,Dixon:2013haa,deFlorian:2013psa,Coradeschi:2015tna,Cieri:2017kpq,Campbell:2017rke},
and then extended beyond NLO in Ref.~\cite{Bargiela:2022dla}. 
In the latter, the authors employ the soft-virtual approximation at NNLO.
In a nutshell, this amounts to including only the contributions arising from loop diagrams and soft real emissions at NNLO. 
The motivation behind this approach lies in the fact that large absorptive contributions originate from loop diagrams which develop imaginary parts. For more details regarding the soft-virtual approximation and its validity at cross-section level, 
we refer the interested reader to Refs.~\cite{deFlorian:2012za,Bargiela:2022dla,Bonvini:2013jha}.

We now report on the main findings of Ref.~\cite{Bargiela:2022dla}, starting with specifying the setup of the calculation. 
The $G_\mu$ scheme for electroweak parameters is employed,
using the following as input
\begin{equation}
    G_F = 1.16639 \cdot 10^{-5} \,{\mathrm{GeV^{-2}}}, \,\,
    m_W = 80.398\, {\mathrm{GeV}},
    \,\,
    m_Z = 91.1876\, {\mathrm{GeV}},
\end{equation}
which give the electromagnetic coupling constant $\alpha = 1/132.338.$ The Higgs mass is set to $m_H = 125 \,{\mathrm{GeV}}$ and the top and bottom quarks mass are set to, respectively, $m_t = 173.2 \,{\mathrm{GeV}}$, $m_b = 4.18 \,{\mathrm{GeV}}$.
The authors of~\cite{Bargiela:2022dla} use a in-house parton-level Monte Carlo generator 
where they employ the \texttt{LHAPDF}
library~\cite{Buckley:2014ana} and \texttt{Hoppet}~\cite{Salam:2008qg,Karlberg:2025hxk}
for various parton distribution functions (PDFs) 
manipulations. 
Predictions are derived using the \texttt{NNPDF31\_nnlo\_as\_0118} PDF set~\cite{NNPDF:2017mvq}. The value of the strong coupling constant, $\alpha_s(m_Z) = 0.118$, is extracted from the PDF set. 
The renormalization and factorization scales are set to $\mu_R=\mu_F=m_{\gamma \gamma}/2$, where $m_{\gamma \gamma}$ denotes the invariant mass of the diphoton system.

\begin{table*}
\renewcommand*{\arraystretch}{1.4}
\centering
\begin{tabular}{cccc}
Order  &  $\sigma_\mathrm{sig}$ [fb] & $\sigma_\mathrm{int}$ [fb] & $\delta_\mathrm{int}$ [\%] \\
\hline
\hline
\multicolumn{4}{c}{$\sqrt{s}=7$~TeV}\\
\hline
LO & $9.04^{+2.10}_{-1.66}$ & $-0.03^{-0.01}_{+0.01}$ & $-0.33$ \\
\hline
NLO & $20.78^{+4.68}_{-3.43}$ & $-0.25^{-0.10}_{+0.07}$ & $-1.22$ \\
\hline
$\mathrm{NNLO}^{*}$ & $26.00^{+2.21}_{-2.49}$ & $-0.45^{-0.06}_{+0.07}$ & $-1.71$ \\
\hline
\hline
\multicolumn{4}{c}{$\sqrt{s}=8$~TeV}\\
\hline
LO & $11.22^{+2.42}_{-1.95}$ & $-0.04^{-0.02}_{+0.01}$ & $-0.39$ \\
\hline
NLO & $26.01^{+5.72}_{-4.18}$ & $-0.32^{-0.11}_{+0.08}$ & $-1.24$ \\
\hline
$\mathrm{NNLO}^{*}$ & $32.82^{+2.93}_{-3.14}$ & $-0.56^{-0.06}_{+0.08}$ & $-1.70$ \\
\hline
\hline
\multicolumn{4}{c}{$\sqrt{s}=13$~TeV}\\
\hline
LO & $23.05^{+3.60}_{-3.20}$ & $-0.13^{-0.04}_{+0.03}$ & $-0.57$ \\
\hline
NLO & $55.18^{+11.08}_{-8.06}$ & $-0.72^{-0.19}_{+0.15}$ & $-1.31$ \\
\hline
$\mathrm{NNLO}^{*}$ & $70.17^{+6.42}_{-6.30}$ & $-1.18^{-0.08}_{+0.12}$ & $-1.69$ \\
\hline
\hline
\multicolumn{4}{c}{$\sqrt{s}=13.6$~TeV}\\
\hline
LO & $24.54^{+3.71}_{-3.33}$ & $-0.14^{-0.04}_{+0.03}$ & $-0.59$ \\
\hline
NLO & $58.92^{+11.72}_{-8.53}$ & $-0.77^{-0.20}_{+0.16}$ & $-1.32$ \\
\hline
$\mathrm{NNLO}^{*}$ & $74.90^{+6.77}_{-6.74}$ & $ -1.27^{-0.08}_{+0.13}$ & $-1.69$ \\
\hline
\hline
\end{tabular}
\caption{Results for the integrated cross section at different values of the collider center-of-mass energy for the diphoton production mode in the setup of Eq.~\eqref{eq:fiducialsetup}. The left column refers to the perturbative order of the calculation, the second one reports the results for the signal cross section only, while the third column includes the contribution from the interference term. The fourth column shows the relative impact of the interference on the signal rate at the nominal scale setting. The asterisk on the NNLO entry indicates that the signal rate is computed in full QCD, whereas the interference term in soft-virtual approximation. For details on the latter and on estimation of theory uncertainties we refer the reader to the main text.}
\label{tab:resultsgammagamma}
\end{table*}
We now present results for four different values of center-of-mass energy in LHC collisions, i.e. $7$, $8$, $13$ and $13.6\,{\mathrm{TeV}}$. We consider a fiducial region defined by the following set of cuts on the final-state photons
\begin{gather}
\label{eq:fiducialsetup}
100 \, \mathrm{GeV}<m_{\gamma\gamma}<150 \, \mathrm{GeV},\quad\;
\quad p_{T,\gamma} > 20\,\GeV, \quad\;
\sqrt{p_{T,\gamma_1} p_{T,\gamma_2}} > 35\,\,\GeV, \nonumber \\
\quad\quad |y_{\gamma}| < 2.5, \quad\;
\Delta R_{\gamma_1\gamma_2} > 0.4 \;.
\end{gather}
The selection criterion on the geometric mean of photon transverse momenta is designed to reduce sensitivity to infrared physics~\cite{Salam:2021tbm}. 
The latter selection cut does not have a significant impact in the current setup since hard radiation is neglected at NNLO. However, in Ref.~\cite{Bargiela:2022dla} it was noted how this setup helps improving the quality of the soft-virtual approximation compared to the exact prediction at NLO QCD.
Also, we stress that we do not
employ any isolation condition on the final-state photons. 
While this is clearly not experimentally sound, for the purpose of the current technical study it does not pose any concern related to IR singularities since the photons originate from the Higgs-boson decay.
Our results for the total cross section in this setup are reported in Table~\ref{tab:resultsgammagamma}. 

The LO results for both the signal and interference terms, $\sigma_\mathrm{sig}$ and $\sigma_\mathrm{int}$ respectively, are obtained by retaining the full top- and bottom- mass dependence in the scattering amplitudes. Moreover, as discussed in Section~\ref{se:interferencevv}, at this order the bottom quark mass is the only source responsible for a non-vanishing absorptive interference. As shown in Table~\ref{tab:resultsgammagamma}, such effect is well below percent level and it can be safely neglected at higher orders. 
For this reason, at NLO and beyond we consider the Higgs production mechanism in the infinite top-mass limit. The Higgs decay into a pair of photons is treated at LO in the full SM. This is justified by the fact that QCD corrections to the decay amplitude are known to be small~\cite{Djouadi:1990aj}.
In the background amplitude, the bottom mass is set to zero beyond LO,
and top quarks contributions are neglected. 

The NNLO row of Table~\ref{tab:resultsgammagamma} deserves some comments. As already discussed, the soft-virtual approximation is employed for the interference contribution to the cross section. Hence the label "NNLO" in table~\ref{tab:resultsgammagamma} should be intended as NNLO soft-virtual when referring to the interference.
In the case of the signal cross section, we provide instead exact NNLO results obtained using the parton level generator \texttt{NNLOJET}~\cite{NNLOJET:2025rno}.

For each entry relative to $\sigma_\mathrm{sig}$ and 
$\sigma_\mathrm{int}$ in Table~\ref{tab:resultsgammagamma} we report the prediction obtained in the nominal scale setting $\mu_R=\mu_F=m_{\gamma\gamma}/2$ and we estimate the theoretical uncertainty by varying both scales simultaneously by a factor 2. The value in the superscript refers to the shift in the central value induced by choosing $\mu_R=\mu_F=m_{\gamma\gamma}/4$, whereas the 
subscript corresponds to $\mu_R=\mu_F=m_{\gamma\gamma}$.

We observe that the impact of the interference contribution is consistent across collider energies, leading to a $\sim -1.7\%$ reduction of the cross section at NNLO QCD accuracy.

Finally, we stress that, since we work in the soft-virtual approximation, our analysis provides a reliable description of interference effects on inclusive observables, such as the integrated cross section and the line shape. As hard radiation effects are neglected, we cannot draw quantitative conclusions for differential observables, such as the transverse momentum of the photon pair or of the individual photons. We defer such studies to future work.
\section{The $H\to Z\gamma$ decay channel}
\label{sec:zgamma}
In this section we report results for interference effects in the $H\to Z\gamma$ decay channel. 
Evidence for this rare decay mode was presented in combination by the ATLAS and CMS collaborations in Ref.~\cite{ATLAS:2023yqk}. This first investigation reported a signal yield $\mu = 2.2 \pm 0.7$, signalling a strong tension with the SM prediction. However, in the most recent ATLAS~\cite{ATLAS:ZAconf} and CMS analyses~\cite{CMS:ZAconf}, this tension has significantly reduced and the reported signal yield is $\mu=1.3^{+0.6}_{-0.5}$ and $\mu=1.10^{+0.52}_{-0.61}$ for ATLAS and CMS respectively, 
thus consistent with the SM expectation. 
From the theory side, the state-of-the-art predictions for this decay channel comprise NLO QCD corrections~\cite{Spira:1991tj,Gehrmann:2015dua} and,
since recently, NLO Electroweak (EW) ones~\cite{Chen:2024vyn,Sang:2024vqk,ReyesR:2025dok}.
QCD corrections have been found to be negligible, whereas EW ones amount to several percents and significantly
improve the theory uncertainty related to the choice of EW input parameters~\cite{Chen:2024vyn,Sang:2024vqk,ReyesR:2025dok}.
As for signal-background interference, this was only investigated at LO accuracy and beyond in Ref.~\cite{Buccioni:2023qnt}. Here we summarize the main outcome of this study.

The setup chosen for the calculation is inspired by an analysis of the $H\to Z\gamma$ decay channel carried out by the ATLAS collaboration~\cite{ATLAS:2020qcv}. We consider the decay of the $Z$ boson into an electron-positron pair and require such pair to have an invariant mass $m_{ll}$ in the range $50\,\mathrm{GeV} < m_{ll} < 101\,\mathrm{GeV}$. Both the two leptons and the tagged photon are required to have a minimum transverse momentum, specifically $p_{T,i} > 15\,\mathrm{GeV},\,\,i\in [e^-,e^+,\gamma]$. Moreover, rapidity selection criteria 
are imposed as $|y_{e^\pm}| < 2.47$ and $|y_\gamma| < 2.37$. 
As in the previous section, we employ the $G_\mu$ scheme with the same value of the relevant input parameters. We use the \texttt{NNPDF31\_nlo\_as\_0118} PDF set~\cite{NNPDF:2017mvq} as distributed through \texttt{LHAPDF} and extract the strong coupling constant from there, i.e.~ $\alpha_s(m_Z) = 0.118$. The QCD renormalization and factorization scales are taken to be $\mu_F = \mu_R = m_{Z\gamma}/2$.

Since we want to quantify the impact of the interference term beyond LO in QCD, we truncate our predictions at NLO accuracy.
As far as Higgs boson production is concerned, 
we work in the infinite top-mass limit. 
Although including top- and bottom-quarks mass effects does not pose any challenge, it was observed in Ref.~\cite{Buccioni:2023qnt} that, similarly to the $H\to\gamma\gamma$ case, such effects are negligible. Indeed, the dominant absorptive contribution to the interference comes from loops of light fermions in the background amplitude, 
which in this case contribute already at LO.
The Higgs decay instead is computed at LO in the full SM.

As for the interference term, the soft-virtual approximation is employed at NLO. Therefore, only contributions originating from loop amplitudes and soft real emissions are fully taken into account. As discussed already, this is motivated by the origin of large absorptive contributions from purely virtual diagrams.
The background two-loop amplitudes were computed in Ref.~\cite{Gehrmann:2013vga} where quarks running in the loop are treated as massless.

\begin{table*}
\renewcommand*{\arraystretch}{1.4}
\centering
\begin{tabular}{cccc}
Order  &  $\sigma_\mathrm{sig}$ [fb] & $\sigma_\mathrm{int}$ [fb] & $\delta_\mathrm{int}$ [\%] \\
\hline
\hline
\multicolumn{4}{c}{$\sqrt{s}=13$~TeV}\\
\hline
LO & $0.44^{+0.07}_{-0.06}$ & $-0.019^{-0.003}_{+0.003}$ & $-4.3$ \\
\hline
$\mathrm{NLO}^{*}$ & $1.03^{+0.20}_{-0.15}$ & $-0.032^{-0.005}_{+0.004}$ & $-3.1$ \\
\hline
\hline
\multicolumn{4}{c}{$\sqrt{s}=13.6$~TeV}\\
\hline
LO & $0.47^{+0.07}_{-0.06}$ & $-0.020^{-0.003}_{+0.003}$ & $-4.3$ \\
\hline
$\mathrm{NLO}^{*}$ & $1.10^{+0.22}_{-0.16}$ & $-0.034^{-0.006}_{+0.005}$ & $-3.1$ \\
\hline
\hline
\end{tabular}
\caption{Results for the integrated cross section at different values of the collider center-of-mass energy in the $Z\gamma$ decay mode. 
They layout is the same as in Table~\ref{tab:resultsgammagamma}. 
The asterisk on NLO indicates that the signal is computed in full QCD, whereas the interference in soft-virtual approximation. For details on the latter, the fiducial setup and theory uncertainties estimation see main text.}
\label{tab:resultszgamma}
\end{table*}
We summarize our results for the 13 TeV and 13.6 TeV LHC in Table~\ref{tab:resultszgamma}. Here
we report the signal rate $\sigma_{\mathrm{sig}}$ alongside the destructive interference $\sigma_{\mathrm{int}}$.
The theory uncertainties reported in the table are estimated by a simultaneous rescaling of the nominal renormalization and factorization scales by a factor $1/2$ and $2$. Superscripts refer to a shift in the central value induced by a factor $1/2$ rescaling, whilst subscripts to a factor $2$ rescaling. 

We immediately observe that at LO the reduction of the signal rate by the interference contribution is very close to the rough estimate in Eq.~\eqref{eq:interferenceestimate}. Also, this is significantly larger than in the diphoton case. As discussed, this is due to the presence of a massive $Z$ boson which provides an imaginary part in the one-loop background amplitude with massless quarks already at LO. This $-4.4\%$ effect is reduced to $-3.2\%$ at NLO. This can be understood because the  $K$-factor of the signal is larger than that of the pure background and of the interference term as well. It would be sensible to expect the same trend to continue at NNLO QCD. However, several ingredients are currently missing in order to address this investigation, most notably three-loop QCD amplitudes for the background process.
%
%
%
\section{Conclusions}
\label{sec:conclusions}
With the LHC continuing to collect data, and even more so in its High-Luminosity phase, controlling the theoretical uncertainty on the Higgs boson production rate in the dominant gluon-fusion channel remains a central task for the theory community. Until now, the uncertainty budget has typically been evaluated on the production side only, treating the Higgs decay mode separately. This approach was well justified as long as the target precision exceeded the percent level. However, multiple sources of uncertainty at the percent level can accumulate and affect the total cross section by several percent.

In this contribution, we highlight the shortcomings of isolating the production from the decay stage. In particular, interference effects in signatures where the Higgs decay is loop-induced can modify the cross section beyond the one-percent level. This is especially relevant for the diphoton signature, which is amongst the most precisely measured Higgs decay channel. Similar, and in fact larger, effects arise in the $Z\gamma$ decay mode. In that case, however, the total uncertainty is still dominated by lack of statistics, making it unlikely that interference effects will become observable in the near future.

The main results of the paper are presented in Table~\ref{tab:resultsgammagamma} for the diphoton channel, and in Table~\ref{tab:resultszgamma} for the $Z\gamma$ one.
They build upon recent advances in higher-order perturbative QCD calculations, notably the computation of multi-loop amplitudes for the relevant background processes. For the 13~TeV and 13.6~TeV LHC, we find that state-of-the-art predictions in the diphoton channel lead to a 1.6\% reduction in the signal rate, while in the rare $Z\gamma$ mode the destructive interference reaches the 3\% level. The impact of these effects can now be consistently incorporated together with other more traditionally investigated sources of uncertainty in the gluon-fusion production mode.
Therefore, we recommend accounting for the destructive interference effect $\delta_\mathrm{int}$, reported in Tables~\ref{tab:resultsgammagamma} and \ref{tab:resultszgamma}, in fiducial cross-section measurements of central Higgs-boson production at the LHC.

\section*{Acknowledgments}
We are grateful to Matteo Marcoli for helping us obtaining predictions for the fiducial Higgs boson cross section at NNLO QCD with \texttt{NNLOJET}.
We also thank Fabrizio Caola and Lorenzo Tancredi for comments on the manuscript.
F.D. is supported by the United States Department of Energy under contract DE-AC02-76SF00515.

\bibliography{biblio.bib}

\end{document}